# Tuning magnonic devices with on-chip permanent micromagnets

M. Cocconcelli, S. Tacchi[*], Róbert Erdélyi, F. Maspero, A. Del Giacco, A. Plaza, O. Koplak, A. Cattoni, R. Silvani, M. Madami, Á. Papp, G. Csaba, Felix Kohl, Björn Heinz, Philipp Pirro, R. Bertacco

*corresponding author

M. Cocconcelli, F. Maspero, A. Del Giacco, A. Plaza, O. Koplak, A. Cattoni, R. Bertacco
*Dipartimento di Fisica -Politecnico di Milano, Via G. Colombo 81, 20133 Milano (Italy)*

Raffaele Silvani, Marco Madami
*Dipartimento di Fisica e Geologia, Università di Perugia, Via A. Pascoli, 06123 Perugia, Italy*

Silvia Tacchi
*Istituto Officina dei Materiali del CNR (CNR-IOM), Unità di Perugia, Via A. Pascoli, 06123 Perugia, Italy*

Róbert Erdélyi, Ádám Papp, György Csaba
*Faculty of Information Technology and Bionics, Pázmány Péter Catholic University, Budapest, Hungary; Jedlik Innovation Liability Company, Budapest, Hungary*

Felix Kohl, Björn Heinz, Philipp Pirro
*Fachbereich Physik and Landesforschungszentrum OPTIMAS, Rheinland-Pfälzische Technische Universität Kaiserslautern-Landau, 67663 Kaiserslautern, Germany*

**Abstract**

One of the most appealing features of magnonics is the easy tunability of spin-waves propagation via external magnetic fields. Usually this requires bulky and power-hungry electromagnets which are not compatible with device miniaturization. Here we propose a different approach, exploiting the stray field from permanent micromagnets integrated on the same chip of a magnonic wave-guide. In our monolithic device, we employ two SmCo square micromagnets (10×10 μm$^2$) flanking a CoFeB conduit at different distances from its axis, to produce a tunable transverse bias field between 7.5 and 3.0 mT in the conduit region between the magnets. Spin waves excited by an antenna just outside the region between the magnets enter a region with a variable higher (lower) effective field when an external bias field is applied parallel (antiparallel) to that from the micromagnets. Consequently, the attenuation length and phase shift of Damon-Eshbach spin waves can be tuned in a wide range by playing with the parallel-antiparallel configuration of the external bias and the distance between SmCo micromagnets and the CoFeB conduit. This work demonstrates the potential of permanent micro-magnets for the realization of low-power, integrated magnonic devices with tunable functionalities.

1. Introduction

Over the past decade, a significant innovation has occurred in radio frequency (RF) designs, leading to the emergence of new communication protocols such as 5G. These protocols demand high-frequency data transmission and robust bandwidth selectivity[1] . The reference technology in this respect is exploiting surface (SAW) or bulk (BAW) acoustic waves, which can cover frequencies up to 10 GHz.[2] For integrated RF components SAW technology is currently used in mobile phones but achieving working frequencies above 5 GHz is still a challenge. Furthermore, typical components like RF filters, time delay and phase shifter devices are poorly tunable so that in communication devices supporting a high number of radio frequency many RF components tuned at specific frequency must be included. There is an emerging need for tunable RF components, but so far realistic solutions have been implemented only in macro-devices, [3,4,5,6] while tunable SAW devices are still under development[7].

Magnonics presents considerable promise in addressing these requirements taking advantage from spin waves (SWs) inherent flexibility to process, encode, and transport RF signals[8,9,10,11]. Indeed, magnons allows for the miniaturization of devices operating in the frequency range between several GHz and THz, thanks to the smaller wavelength of SWs as compared to electromagnetic waves oscillating at the same frequency. Furthermore, using the SWs phase provides an additional degree of freedom in data processing, while the non-linear phenomena can be exploited to implement unconventional functionalities.[12,13] These features are extremely appealing both for RF signal processing and wave-computing applications. [14] Finally, magnonics based on 3d alloys has shown good compatibility with traditional complementary metal–oxide–semiconductor (CMOS) technology [15].

However, the practical implementation of magnon technology in real devices encounters limitations arising from challenges in integrating magnonic circuits with conventional electronics. This is due to the necessity of a bias magnetic field, essential for tuning the spin wave propagation. Usually power-hungry and space-consuming approaches are used, where the required field is provided by a current sent through a lithographically defined wire[16,17] or external electromagnets which cannot be seamlessly integrated into consumer devices.[18,19,20,21]

A potential solution to these constraints involves incorporating the source of the magnetic field on the chip. Integrating permanent micromagnets (µ-magnets) on the same chip as the magnonic waveguide offers the possibility of creating fully integrated, self-standing magnonic devices. Such devices could be tunable by varying the distance between the waveguide and the µ-magnets, a feature that could become achievable implementing micro-electromechanical systems to vary the proximity of the magnets to the waveguide.

In the path towards this high level of integration, it is crucial first to demonstrate the feasibility of leveraging the µ-magnet proximity to manipulate SW propagation. In this study, we present a first example of on-chip integration of permanent µ-magnets and magnonic structures.

The functionality of our devices, consisting of a CoFeB SW conduit, placed between two square SmCo permanent µ-magnets, was evaluated by means of micro-focused Brillouin Light Scattering (micro-BLS). From the measurements of the thermal SW signal we find that, changing the distance between two (10x10 µm) SmCo µ-magnets flanking a CoFeB conduit (1.8 µm wide), the local field by the µ-magnets in the central region of the conduit can be tuned in the range between 3.0 and 7.3 mT. To gain a deep physical understanding of the impact of the field from the µ-magnets on SWs in the CoFeB conduit, we carefully investigated Damon-Eshbach modes excited at a fixed wave-vector, for different distances between the µ-magnets and the conduit. The experiments were performed applying a uniform external magnetic field along the short axis of the CoFeB conduit, either parallel or antiparallel to the field produced by the µ-magnets. We find that the proximity of the µ-magnets is not critical for SW excitation and propagation, due to the large magnetic anisotropy of SmCo which prevents an effective dynamic coupling to the antenna and the CoFeB conduit. On the other hand, the variation of the magnetic field along the conduit, due to the presence of the local stray field of the µ-magnets, deeply affects the SW propagation. In particular, for the parallel alignment a strong attenuation of SWs is observed, together with a sizable dependence of the decay length on the µ-magnet-conduit distance.

To estimate the potential of this device layout to tune SW propagation we simulated the behavior at fixed excitation frequency, for both parallel and antiparallel alignment between the external field and that from the µ-magnets on changing the magnet-conduit distances. Large modulations of the decay length (3-10 µm) and phase shift (± 5 rad) are found, thus pointing out the huge potential of integrated permanent magnets for the implementation of tunable magnonic devices.

2. Results and discussion

2.1 Device concept and realization

To significantly influence the propagation of spin waves (SWs), our objective is to generate a local magnetic field at the conduit position, using µ-magnets as integrated magnetic field sources. To achieve this, we employ a hard magnetic material (SmCo) characterized by sizable remanent magnetization and a coercivity sufficient to prevent magnetization switching under fields with magnitude needed to bias the magnonic conduit in the Damon Eshbach (DE) configuration.

The device layout, depicted in Figures 1a,b comprises two square 10×10 µm SmCo permanent µ-magnets[22] on a silicon substrate, sandwiching a CoFeB conduit for SW propagation along the y axis. W(100 nm)/SmCo(500 nm)/W(100 nm) multilayers are grown by sputtering on Si(001) substrates. The hard-magnet behavior is achieved by annealing at 650°C, leading to a polycrystalline film containing $SmCo_5$ and $Sm_2Co_{17}$ hard phases. Finally, patterning is performed by optical lithography and ion beam etching (IBE), as described in detail in the experimental section and in Ref [23]. The patterned square µ-magnets exhibit an in-plane remanent magnetization of 0.48 T along the transverse x direction, as measured by VSM upon application of a maximum field of 2 T (see supplementary information S1 and Ref. [22]), and a coercive field on the order of 0.5 T. Notably, prolonged IBE for such thick layers creates a sizable roughness on the Si substrate (60 nm rms), thus requiring a subsequent planarization step prior to the fabrication of the CoFeB conduit and antenna. The µ-magnets are symmetrically positioned with respect to the conduit and intentionally smaller than the conduit length, to provide a symmetric field profile along the x direction, confined in a region along the longitudinal y direction close to the antenna region. This configuration minimizes the impact of the µ-magnet proximity on the SW excitation, while still producing a sizable field in the portion of the conduit within the propagation length of SWs in CoFeB.

Simulations carried out by Comsol Multiphysics indicate that the magnetic field generated by the μ-magnets (Figures 1c) is predominantly directed along the transverse direction and confined in the region between the two μ-magnets, with a rapid decay away from them.

A magnonic waveguide consisting of Ta(5 nm)/CoFeB(25 nm)/Ta(5 nm), 1.8 μm wide and 50 μm long, is positioned between the μ-magnets, as depicted in Figure 1a, b. A planarization layer, composed of silica sol-gel, is utilized both to planarize the Si substrate after IBE and to position the conduit on a plane parallel to the substrate and placed at half height of the μ-magnets, where the component $H_{Mx}$ of the stray field from the μ-magnet ($H_M$) along the magnetization direction (x) is prominent, while the $H_z$ component is negligible. In the plane of the CoFeB conduit the longitudinal profile of $H_M$ (Figure 1d) displays a bell-shape with FWHM roughly corresponding to the μ-magnet size (10 μm), while the transverse profile is quite flat (Figure 1e). A sizable dependence of $H_M$ on the distance D between the conduit axis and the μ-magnet is found. Different devices with variable distance D (5.5, 7.5, 9.5 μm) were fabricated on the same substrate, to implement experimental configurations with variable maximum field intensity (from 7.5 to 3.0 mT) as depicted in Figures 1d,e.

To investigate the magnetic properties of the conduit, we carried out a comprehensive analysis of as-grown CoFeB films, by vibrating sample magnetometry (VSM) measurements and ferromagnetic resonance (FMR). Despite the absence of an applied magnetic field during deposition and the rotation of the sample holder in the confocal sputtering machine, a clear in-plane uniaxial magnetic anisotropy is found in all films. This is evident from Figure S2 in the Supporting Information, reporting the polar plots of the ratio Mr/Ms between the remanence and saturation magnetization measured by VSM as a function of the angle between the applied field and the [110] direction of the Si substrate. To assess whether this anisotropy stems from magnetostriction arising from the substrate, we investigated Ta(5 nm)/CoFeB(25 nm)/Ta(5 nm) films on both bare Si (100) substrates and planarized Si substrates by sol-gel silica layers with variable thickness spun onto silicon (100). A comprehensive analysis of VSM and FMR data for different silica thickness is reported in the Supplementary. The 350 nm silica layer induces a rotation of the uniaxial anisotropy axis by about 20° while we observed no changes in the Gilbert damping ($α_G=5 \times 10^{-3}$) and saturation magnetization between the two cases. The change of the anisotropy direction is attributed to variations in the strain experienced by the CoFeB film when grown on silicon (Si) versus spin-on-silicon dioxide ($SiO_2$), which is a softer material. We suggest that the effect of the substrate on the CoFeB anisotropy can be even more pronounced when comparing conduits fabricated on bare silicon (Si) and on the silica planarization layer between the two micromagnets in the final device (see Figure 1b). The hollow formed between the two magnets induces a saddle-shaped topography in the silica layer, which is expected to impose an additional strain on the CoFeB conduit. We suggest that this is the cause of the relatively large anisotropy constant that we must use to fit BLS data taken on conduits within the magnets in simulations (see section 2.3 for details).

A coplanar waveguide (CPW) was fabricated on top of the sample to excite SWs in the CoFeB conduit. The CPW is placed at 1.5 μm distance from the magnets, to ensure that SWs could reach the μ-magnets region before attenuating below the micro-BLS sensitivity. However, as depicted in Figure 1c, d, this results in the magnet stray field not being completely extinguished at the antenna position.

The local field produced by the μ-magnets deeply impacts on SWs propagation in the DE geometry i.e. for SWs wave vectors perpendicular to the magnetic field direction. The physical behavior of our device can be understood observing that, when an external uniform magnetic field $H_{ext}$ is applied along the short axis of the conduit, $H_M$ induces a local enhancement or reduction of the total magnetic field in the region comprised between the μ-magnets, depending on the relative orientation between $H_M$ and $H_{ext}$. This implies a non-uniform "shift" of the local SW dispersion relation in the different parts of the conduit which affects the SW propagation. In Figure 1f we show the simulated dispersion relations for the DE modes in a conduit under a uniform transverse magnetic field corresponding to that at the center of the CoFeB conduit in three characteristic conditions: when only the external field $H_{ext}$ is present (D→ ∞) and for both a parallel and an antiparallel alignment between $H_M$ and $H_{ext}$. In all the simulations we set $\mu_0 H_{ext}$= 60 mT as in the BLS experiments, whereas $\mu_0 H_M$ was fixed to 7.0 mT, corresponding to the value estimated for the sample having D=5.5 μm. The simulated dispersion curves show the existence of two main branches, corresponding to the fundamental mode of the conduit (continuous line), and the first width quantized mode (dashed line). For k≅ 0.08 2π/μm, corresponding to the wavevector mainly excited by the CPW, the two modes are very close in frequency and beating of the mode profiles can be observed in the simulations (see Supplementary Information for details). Overall, for the parallel (antiparallel) alignment the SWs dispersion curve in the region of the conduit flanked by the μ-magnets is shifted by about 0.5 GHz toward higher (lower) frequencies with respect to the case where only the external field is present. This leads to two different regimes of SWs propagation.

For the parallel configuration, when the excited SWs enter the region between the μ-magnets, where the total magnetic field is larger, the SW dispersion relation is shifted towards higher frequencies. As a consequence, if the frequency of the excited modes is lower than the FMR value in this region, the SW modes are not allowed to propagate in this part of the conduit and a strong attenuation is expected. On the contrary, for the antiparallel configuration, when the excited SWs enter the region between the permanent µ-magnets, where the total magnetic field is lower, the dispersion relation undergoes a shift towards lower frequencies. The frequency of the excited modes always remains above the FMR frequency and a larger propagation distance is expected. On top of this, one must also consider that the shift of the dispersion curves also implies a change of the group velocity at fixed frequency. Based on these considerations, it is easily understood that the proposed device can behave like an attenuator or phase-shifter whose features can be tuned by varying the distance D between the conduit and the µ-magnet.

2.2 BLS experiments

The functionality of the proposed devices was tested by means of micro-BLS, investigating SWs propagation in DE geometry.
In order to quantitatively estimate the magnetic field $\mu_0 H_M$, we first measured the intensity of thermal SWs at the center of the conduit as a function of the distance from the CPW, with an external uniform magnetic field $\mu_0 H_{ext}$= 60 mT applied along the short axis of the conduit. The direction of $H_{ext}$ was set to be either antiparallel or parallel with respect to the stray field produced by the µ-magnets, $H_M$, which is mainly directed perpendicular to the major axis of the conduit.
Figure 2 shows the intensity of thermal SWs plotted versus the frequency shift of the BLS signal and the distance from the CPW, measured for the antiparallel (panel a) and parallel (panel b) alignment. For the antiparallel (parallel) configuration, in the region between the µ-magnets (black dashed lines) the SW frequency at which we observe the onset of the BLS signal (roughly corresponding to the FMR frequency) is found to decrease (increase) with respect to the value measured in the region far from the antenna, where the effect of the field produced by the µ-magnets is negligible. In particular, the FMR frequency variation $\Delta f$=FMR$_P$-FMR$_{AP}$, measured in the region between the µ-magnets, when switching from the parallel to the antiparallel configuration, was used to quantitatively estimate the dependence of $\mu_0 H_M$ on the µ-magnet-conduit distance. $\Delta f$ is found to decrease from about 1.0 GHz for the sample having D=5.5 µm to about 0.4 GHz for D=9.5 µm, indicating a reduction of $\mu_0 H_{mag}$. Exploiting the Kittel equation and fixing the saturation magnetization to the value Ms=1.2·10$^6$ A/m, measured for the continuous film (see Supplementary Materials), we find that $\mu_0H_{mag}$ decreases from about 7.5 mT for D=5.5 µm to about 3.0 mT for D=9.5 µm (see Table I), in very good agreement with the value obtained in the COMSOL simulations (see Fig.1).

Then the propagation of SWs excited by the CPW was studied in the DE geometry for both parallel and antiparallel configurations between $H_{ext}$ and $H_M$, applying an external field $\mu_0H_{ext}$=60 mT. As first step to identify the modes excited by the CPW in the conduit, micro-BLS measurements were performed measuring the SW intensity as a function of the excitation frequency in the range between 2 and 15 GHz, in close proximity to the antenna (about 1.0 µm distance) and before entering the region flanked by the µ-magnets (See Section S6 in supporting Information). In agreement with the structure of the band dispersion, simulated for a uniformly magnetized conduit (see Figure 1f), for all samples the BLS spectra show two peaks corresponding to the fundamental and first width-quantized mode at high and low frequency, respectively. They are found at about 9.0±0.2 GHz and 8.7±0.2 GHz in the antiparallel configuration, and at about 9.6±0.2 GHz and 9.0±0.2 GHz in the parallel one. The small frequency shift observed between the parallel and antiparallel configurations can be ascribed to the field produced by the µ-magnets, which is small but not entirely negligible at the antenna position. Then, two-dimensional maps of the SWs intensity for the two observed modes were acquired at fixed frequency, selecting the values which maximize the excitation efficiency close to the antenna for each experimental configuration (parallel/antiparallel and at variable D). As detailed above, these frequencies can be slightly different, but this experimental protocol ensures that all the modes are excited at a fixed wave-vector, namely that corresponding to the first maximum of the antenna excitation efficiency (k=0.08 2π/µm, as reported in section S4 of the Supporting Information).

Micro-BLS measurements performed for the sample with D=5.5 µm, in the antiparallel and parallel configuration, are reported in Figure 2 (c) and (e), respectively. The maps are recorded as a function of the distance from the antenna over an area of about 3.5×16 µm$^2$ with a 250 nm step size. As it can be seen both the modes are characterized by an almost uniform spatial profile across the width of the conduit, due to the hybridization of the two SWs modes, in agreement with micromagnetic simulations.

In the antiparallel configuration, the mode at higher frequency is observed up to about 16 µm distance from the antenna, while a smaller propagation distance is observed for the mode at lower frequency (Figure 2 (c)). When the external field is switched in the parallel configuration, instead, the propagation of both the modes is drastically attenuated (Figure 2(e)). In particular, for the mode at lower frequency, a rapid reduction of the micro-BLS signal can be observed within a few microns from the CPW. Moreover, one can observe that when entering in the region between the µ-magnets, the mode transforms in two beams propagating in the edge regions of the conduit. The decay length for the SW intensity has been estimated from the fit of the micro-BLS intensity profile taken along the y direction at the center of the conduit, as shown in Fig.2 (d) and (f), using the equation $I(y) = I_1 \exp\left(-\frac{2y}{\lambda_D}\right) + I_0$, where $y$ is the position along the waveguide, $I_1$ the SW intensity at the antenna position, and $I_0$ the offset baseline due to noise and detector dark count effects. In the antiparallel configuration a decay length of about 4.6 µm and 3.8 µm has been found for the high and low frequency mode, respectively. Note that, in this configuration, SWs propagate efficiently over a distance comparable to that previously reported for CoFeB waveguides[24,25] suggesting that the proximity of the permanent µ-magnets is not draining power due to spin-waves excitation. This also indicates that, although the local variation of the magnetic field induces a shift of the dispersion curve and consequently a variation of the wave vector of the SW modes, a significant reflection is not observed thanks to the backscattering protection of the DE modes.[26]

In the parallel configuration, $\lambda_D$ exhibits a noticeable reduction, with a decrease of about 30% (60%) for the high (low) frequency mode. The strong attenuation of the low frequency mode and its splitting in two beams can be explained observing that the frequency of this mode (8.85 GHz) is smaller than the FMR value in the region in between the µ-magnets (about 9.1 GHz from Figure 2b). Therefore, the mode is not allowed to propagate in this part of the conduit and localizes in the edge region where the internal field is reduced due to the demagnetizing effects.
[27]On the contrary, the mode at higher frequency (9.55 GHz) remains above the FMR value in the region between the µ-magnets and can propagate, but a marked reduction of $\lambda_D$ is found. This is because in the region between the µ-magnet, due to the shift at higher frequency of the dispersion relation, the same frequency is supported by the CoFeB conduit for smaller wavevector where the SWs group velocity is reduced.

Finally, we investigated the effect of the distance between the µ-magnets and the conduit on the propagation properties. Figure 3 shows a comparison of the two-dimensional maps of BLS intensity for D = 5.5, 7.5, and 9.5 µm, measured in the antiparallel (parallel) configuration for both the low and high frequency mode. In the antiparallel configuration, both the high- and low-frequency modes exhibit similar behaviour for all the investigated samples whit a decay length of about 4.6-4.8 µm and 3.8-3.4 µm, respectively (see Table I). On the contrary, in the parallel configuration, the attenuation of the SWs propagation and the change of $\lambda_D$ become gradually less pronounced on increasing D. This can be explained considering that when D increases $\mu_0 H_M$ decreases and, correspondingly, the shift of the dispersion relation towards higher frequency becomes smaller.

For the sample with D=9.5 μm, the values of $\lambda_D$ for both the parallel and antiparallel configuration are very similar, in agreement with the fact that, at this distance, the variation of the local field induced by the μ-magnets (3.0 mT over 60 mT of $H_{ext}$) only marginally affects SWs propagation.

| D (μm) | $\mu_0 H_M$ (mT) | antiparallel configuration | | parallel configuration | |
|---|---|---|---|---|---|
| | | Mode freq. (GHz) | $\lambda_D$ (μm) | Mode freq. (GHz) | $\lambda_D$ (μm) |
| 5.5 μm | 7 | 9.0 / 8.6 | 4.6±0.2 / 3.8±0.1 | 9.55 / 8.85 | 3.1±0.1 / 1.6±0.1 |
| 7.5 μm | 4 | 9.30 / 8.80 | 4.8±0.2 / 3.8±0.1 | 9.70 / 9.15 | 4.0±0.1 / 2.0±0.1 |
| 9.5 μm | 3 | 9.25 / 8.75 | 4.8±0.1 / 3.4±0.1 | 9.55 / 9.05 | 4.6±0.1 / 3.8±0.1 |

**Table I:** Summary of the parameters obtained via micro-BLS analysis on devices with different distance (D) between the magnets and the conduit. $\mu_0 H_M$ is the intensity of the stray field produced by the permanent μmagnets, estimated from the measurements of the thermal signal. The frequency of the main SWs mode excited by means of the CPW and the corresponding decay length ($\lambda_D$), obtained from the analysis of the BLS intensity, are reported for the antiparallel and parallel configurations.

## 2.3 Discussion

BLS measurements show that our device behaves like a "spin-wave valve", where the transmission of SW modes is inhibited in the parallel configuration, while it is enabled in the antiparallel configuration. In the parallel configuration, corresponding to the "off state" of the spin-wave valve, the attenuation of both modes can be largely tuned by changing the distance between the μ-magnets and the conduit. We can estimate the dependency of the attenuation on the distance D with reference to a possible device with a detector of transmitted SWs placed at a distance y=14 μm from the antenna. Neglecting the insertion losses due to the input and output devices (transmitter and receiver) which could be implemented by antennas or other devices, we can estimate the attenuation in dB in the case of a distance D = 5.5, 7.5 μm with respect to the reference case of Dr = 9.5 μm, where the impact of the magnet stray field is negligible.

Using the formula $A(D) = 10 \log_{10}\left(\frac{I_D}{I_{Dr}}\right) = 10 \log_{10}\left(\frac{e^{-2y/\lambda_D}}{e^{-2y/\lambda_{Dr}}}\right)$ we find that, in the parallel configuration, approaching the magnets from 9.5 μm to 7.5 μm (5.5 μm) introduces an additional attenuation of -16.8 dB (-64.2 dB) for the high frequency mode. For the low-frequency mode the corresponding attenuations are -11.0 dB and -24.0 dB.

To gain a deeper understanding on the physical behavior of our device and to visualize also the effects on the SW phase, we performed micromagnetic simulations including SW excitation by the real antenna used in the experiments. The geometry of the simulated device, corresponding to the image of the real device in Figure 1, is reported in the Supplementary Information (Figure S6). The stray field from the μ-magnets used in dynamic simulations has been evaluated with Mumax and it is fully consistent with that reported in Figure 1, obtained with COMSOL. Details on the micromagnetic parameters used in the simulation are reported in the methods.

To investigate the potential of the proposed device for manipulating SW propagation, we simulated its functionality at a fixed excitation frequency. Note that this is a different approach with respect to the BLS analysis, where we studied the propagation of the SWs modes at the peculiar frequency which maximizes the excitation efficiency in the region of the antenna. In this way the modes are investigated at fixed wavevector, corresponding to the peak of the excitation efficiency of the CPW. This allows to focus on the effect of the μ-magnets on the propagation properties and to minimize possible differences in the excitation efficiencies arising from small variations of the geometrical parameters in different devices. In the simulations reported below, instead, we select the frequency at which the intensity of the high-frequency mode is maximum when only $\mu_0 H_{ext}$=60 mT is applied, corresponding to the absence of μ-magnets (D → ∞). That frequency (9.25 GHz) is kept fixed in all simulations carried out in the parallel/antiparallel configuration and for the different values of the magnet-conduit distance D. In this way we can simulate the behavior of a single device operated with a monochromatic RF input, for different values of the distance D and configurations, which impact both on SW excitation and propagation to determine the overall device functionality.

Fig. 5a shows the SWs intensity profiles (modulus of the reduced magnetization - left panels) and waveforms (out-of-plane component of the reduced magnetization - right panels) simulated for the three configurations: when only the external field $\mu_0 H_{ext}$=60 mT is applied along the short axis of the conduit and for both a parallel and an antiparallel alignment between the $H_M$ produced by μ-magnets at D=5.5 μm and $H_{ext}$.

In agreement with the experimental results, a drastic decrease of the propagation distance is observed for the parallel configuration. Fig. 4c reports the amplitude decay length ($\lambda_D$) obtained for the simulated intensity patterns as a function of distance in the parallel and antiparallel configurations. $\lambda_D$ was estimated from the fit of the intensity profile taken along the propagation direction at the center of the conduit, using the equation $I(y) = I_0 e^{-\frac{2y}{\lambda_D}}$, where y is the position along the waveguide, and $I_0$ the SW intensity at the antenna position. One can observe that for the parallel configuration $\lambda_D$ shows a very strong decrease on reducing the distance D. In this configuration, indeed, after the excitation SWs enter the region between the μ-magnets, where the total magnetic field is larger and SWs dispersion relation is shifted towards higher frequencies and only the mode at lower frequency can exists at a frequency of 9.25 GHz (See Figure 1f). This causes a transition from the high-frequency mode excited close to the antenna, where the total field is close to $H_{ext}$ so that 9.25 GHz is essentially the frequency of the high-intensity fundamental mode, to the lower frequency mode which is characterized by a lower group velocity and consequently by a lower decay length.

On the contrary for the antiparallel alignment one can observe an increase of $\lambda_D$ on decreasing D. Since at the CPW position the μ-magnet stray field is not completely extinguished and in the antiparallel configuration the dispersion relation is shifted towards lower frequencies with respect to case where the μ-magnets are not present, for small D (large $H_M$) the same frequency is excited at larger wavevector where the SWs group velocity is bigger. Notice that a constant value of $\lambda_D$ is found in the experimental results. This difference can be explained taking into account the different approach used in BLS experiments and in the micromagnetic simulations, as discussed above. Overall, larger values of the decay length are found in the micromagnetic simulations with respect to the BLS analysis, but this can be ascribed to the unavoidable imperfections and roughness of the real samples which are not included in the micromagnetic simulations.

Finally, the phase shift induced by the μ-magnets has been estimated versus D in the parallel and antiparallel alignment. To this scope, temporal recording of the out-of-plane component of the reduced magnetization was done in two distinct points (A and B) along the centerline of the conduit, 1 μm and 13 μm away from the antenna, i.e. immediately before and after the region flanked by the μ-magnets (see Fig. S6 in the Supplementary Information). Fig. 4b illustrates the method for evaluating the phase difference between the two points in a representative configuration (antiparallel, where the SmCo magnets are placed 5.5 μm from the centerline of the conduit). The phase shift that as a function of the magnet-conduit distance in the parallel and antiparallel configurations was calculated with reference to the scenario without magnets (zero phase shift). Depending on the distance D, the wavelength in the central region between the μ-magnets is varied, resulting in an adjustable time delay/phase delay through the region. When D decreases and consequently $H_M$ increases, the phase shift increases up to a value of 5 rad, with an opposite sign in the two configurations, as reported in Figure 4c. This indicates that the device can be operated as a time-delay unit, or phase shifter, by changing the distance between the μ-magnets and the conduit and reversing the alignment between the external applied field and the stray field produced by the μ-magnets. According to Figure 4c, where both the attenuation length and the phase shift are reported for the different configurations, the device could be used as tunable attenuator or phase-shifter at fixed excitation frequency, provided that the distance D can be changed, for instance using a MEMS actuator. The same device could also behave like a tunable filter in the parallel configuration, with a cut-off frequency corresponding to the FMR value at the center of the conduit determined by the intensity of the local field produced by the μ-magnets. Even though the accurate determination of the functional features of real devices based on our concept is beyond the scope of this paper, our work clearly shows the potential of the usage of on-chip micromagnets for magnonics.

3. Conclusion

To summarize, in this work we have experimentally demonstrated the efficient on-chip integration of permanent μ-magnets with magnonic conduits. Using micro-focused BLS we performed a comprehensive investigation of the properties of SWs excited in DE geometry, when an external uniform field is applied along the short axis of the conduit. We demonstrate that the SW propagation in the CoFeB is significantly affected by the local field produced by the μ-magnets. In particular, a significant reduction of the decay length of the propagating SWs is achieved switching from an antiparallel to a parallel alignment between the local stray field, produced by the μ-magnets, and the external one. Moreover, our results show that this effect can be tuned on changing the distance between the conduit and the μ-magnets. Finally, exploiting micromagnetic simulations we show the potential of the proposed devices in terms of the modulation of both the decay length and phase shift as a function of magnets-conduit distance. This work demonstrates that the on-chip integration of permanent micromagnets and magnonic elements is a very promising route towards the implementation of tunable magnonic devices whose functionalities depends on the peculiar local magnetic field landscape.

4. Methods

Sample preparation

A W(100 nm)/SmCo(500 nm)/W(100 nm) stack was deposited onto a silicon substrate using the Leybold LH Z400 sputtering system. Subsequently, the samples underwent annealing in vacuum at 650°C for 30 minutes. Next, 10x10 μm micromagnets were fabricated via optical lithography, using a Heidelberg MLA100 Maskless Aligner, followed by ion beam etching at 300 V beam energy.

A planarization layer of 350 nm $SiO_2$ solgel was then spin-coated onto the sample and baked at 150°C, reducing surface roughness to 0.2 nm/rms while preserving the overall device topography.

Subsequently, a Ta(5 nm)/CoFeB(25 nm)/Ta(5 nm) layer was fabricated between the magnets using optical lithography with the Heidelberg MLA100 Maskless Aligner and lift-off techniques. The deposition process was conducted using DC magnetron sputtering employing an AJA Orion8 system with a base pressure below $1 \times 10^{-8}$ Torr, equipped with a rotating stage. Finally, coplanar GSG antennas were fabricated via optical lithography and lift-off processes, following the deposition of a 25 nm thick SiO2 insulating layer using magnetron sputtering.

***Brillouin Light Scattering***

Micro-BLS measurements were performed by focusing a single-mode solid-state laser (operating at a spectral line of 532 nm) at normal incidence onto the sample using an objective with numerical aperture of 0.75, giving a spatial resolution of about 250 nm. The inelastically scattered light was analyzed by means of a (3+3)-pass tandem Fabry-Perot interferometer. A nanopositioning stage allowed us to position the sample with a precision down to 10 nm on all three axes, and to perform spatial resolved scans moving the sample with respect to the objective. A spatially uniform magnetic field $\mu_0 H_{ext}$ =60 mT, provided by an electromagnet, was applied in the sample plane along the short axis of the conduit. The direction $\mu_0 H_{ext}$ was reversed to be antiparallel or parallel with respect to the direction of the local magnetic field produced by the micromagnets. A DC/AC electrical probe station ranging from DC up to 20 GHz was used for spin-wave excitation, setting the microwave power at +13 dBm on the RF generator output.

***Simulations***

Micromagnetic simulations were carried out by mumax3[Vansteenkiste2014]. Material parameters for CoFeB were extracted from the experimental investigations, $M_{sat}$ = 1229 kA/m, $A_{ex}$ = 14.3 pJ/m, and damping coefficient $\alpha$ = 5.11x10$^{-3}$. The 25 nm thick CoFeB film was simulated as a single layer with 100 nm lateral cell size. In order to obtain a good agreement with the experimental results we set an in-plane uniaxial anisotropy to the value $K_{u1}$ = 12.5

kJ·m$^{-3}$. The source of this anisotropy was traced back to the planarization layer; for the full analysis the reader is referred to the Supplementary materials. The bias field was set to 60 mT in-plane, perpendicular to the spin-wave propagation direction (DE configuration). The SmCo magnets were modeled as static components, the stray fields of the permanent magnets were applied in the dynamic simulations of the spin waves.

We selected the frequency where excitation was the strongest for 60 mT of bias field ($H_{ext}$) in absence of the field $H_M$ by the SmCo micromagnets: f = 9.25 GHz was used, which is close to the experimental measurement frequencies.

For the simulation of the dispersion curve, we (i) used the same geometry (without the SmCo magnets), (ii) applied a Sinc excitation, (iii) recorded the spatio-temporal evolution of the magnetization, and (iv) used 2D Fourier transformation to calculate the dispersion plot. We filtered the data with a gaussian profile with FWHM = 400 nm. This way we can model the filtering effect of the laser spot-size of the BLS in the experiments, omitting the experimentally inaccessible modes from the spectrum.

For the simulation of the whole device at fixed frequency (Figure 4), the SW intensity was calculated by recording the out-of-plane component of the reduced magnetization at 5 ps intervals, starting from the onset of spin wave excitation, and subsequently, these patterns were squared and then averaged over the time range of t = 45-50 ns. The waveforms represent a snapshot of the out-of-plane component of the reduced magnetization at t=50 ns.

Acknowledgements

All the authors acknowledge funds from the European Union via the Horizon Europe project "MandMEMS", grant 101070536. M. M., R. S. and S.T. acknowledge financial support from NextGenerationEU National Innovation Ecosystem grant ECS00000041–VITALITY (CUP B43C22000470005 and CUP J97G22000170005), under the Italian Ministry of University and Research (MUR). R. B. acknowledges funding from NextGenerationEU, PNRR MUR – M4C2 – Investimento 3.1, project IR_0000015 – "Nano

Foundries and Fine Analysis – Digital Infrastructure (NFFA–DI)", CUP B53C22004310006. This work has been partially performed at Polifab, the micro and nanofabrication facility of Politecnico di Milano.

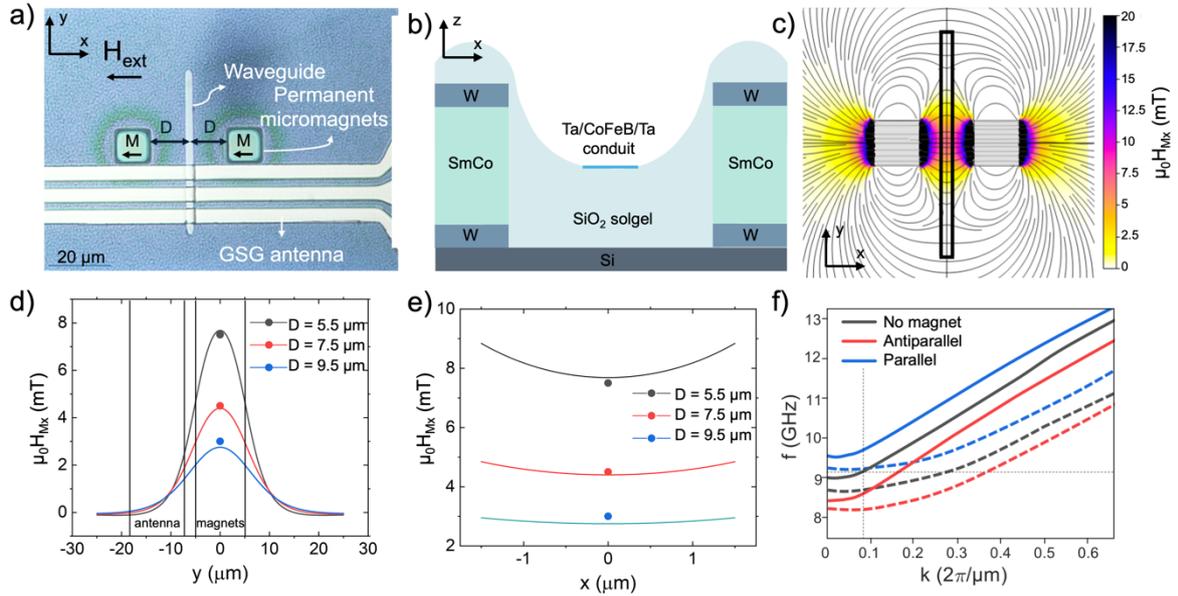

**Figure 1.** a) Optical image of the device and b) cross-section schematics. c) Field lines and color plot of the x component (parallel to the micromagnet magnetization) of the magnetic field $H_M$ generated by the SmCo micromagnets in the plane of the CoFeB conduit. d,e) Simulated (continuous line) and experimental values from BLS (dots) of $H_m$ at different distances (d) from the conduit, along the SW propagation and transverse directions (x and y in panel a); f) Dispersion curves of the CoFeB conduit in DE geometry for a uniform applied transverse field corresponding to three different configurations: when only the external field $\mu_0 H_{ext}=60$ mT is applied and for both the parallel and antiparallel alignment between $\mu_0 H_{ext}$ and $\mu_0 H_M= 7.3$ mT (D = 5.5 μm ). The vertical and horizontal dotted lines indicate the excitation wavevector dictated by the CPW geometry and the fixed excitation frequency of 9.25 GHz used in the simulation of the whole device.

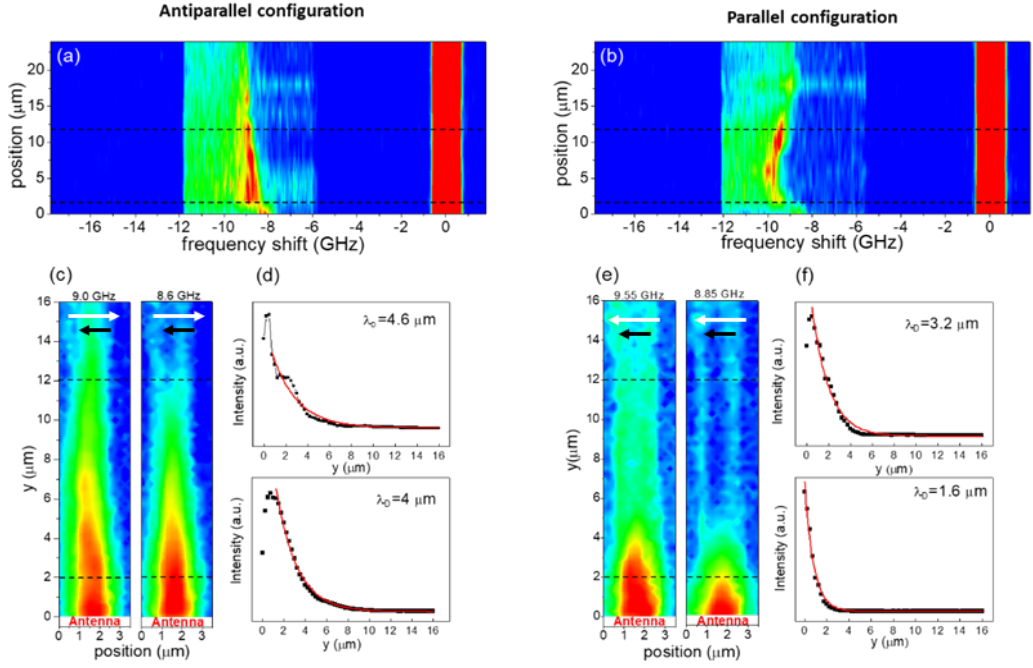

**Figure 2** BLS Analysis of the sample having having a micromagnet-axis distance D of 5.5 μm, in the antiparallel (left panels) and parallel configuration (right panels). (a) and (b) micro-BLS measurements of the thermal SWs signal as a function of the distance from the CPW. (c) and (e) two-dimensional maps of the BLS intensity recorded at the frequencies of the two modes excited by means of the CPW; on the y axis is reported the distance from the antenna along the propagation direction. White and black arrows indicate the direction of the $H_M$ and Hext, respectively. (d) and (f) spin-wave intensity (linear scale) taken at the center of the conduit as a function of the distance y from the antenna (black points) for the higher (top panel) and lower (bottom panel) frequency modes. Black points correspond to experimental data while red lines show the exponential fit with the value of $\lambda_D$ given in the figure.

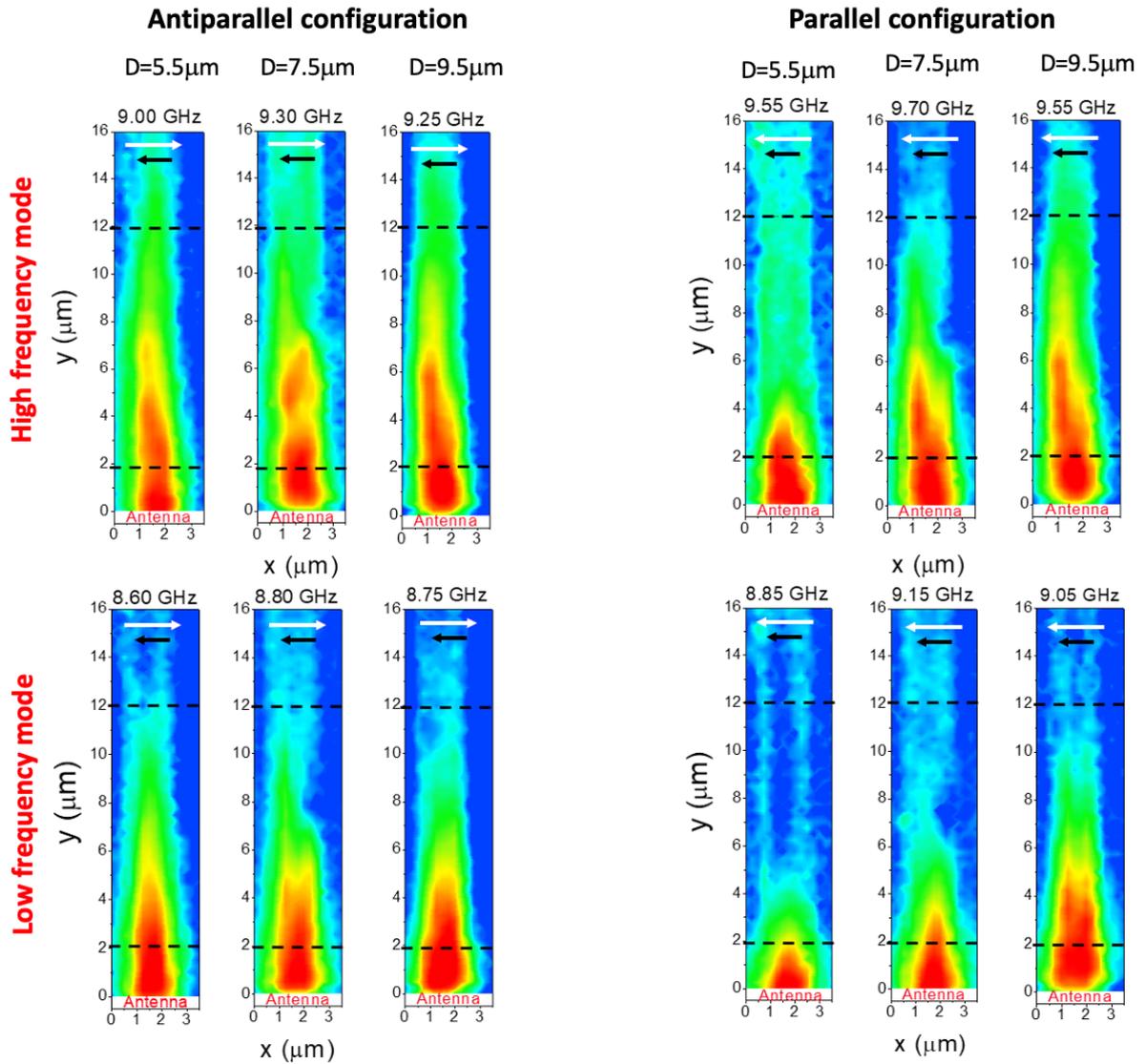

**Figure 3** Two-dimensional maps of the BLS intensity recorded at the frequencies of the two modes excited by means of the CPW in the antiparallel (left panels) and parallel configuration (righ panels) for the high- and low-frequency modes; on the y axis is reported the distance from the antenna along the propagation direction. White and black arrows indicate the direction of the $H_M$ and Hext, respectively.

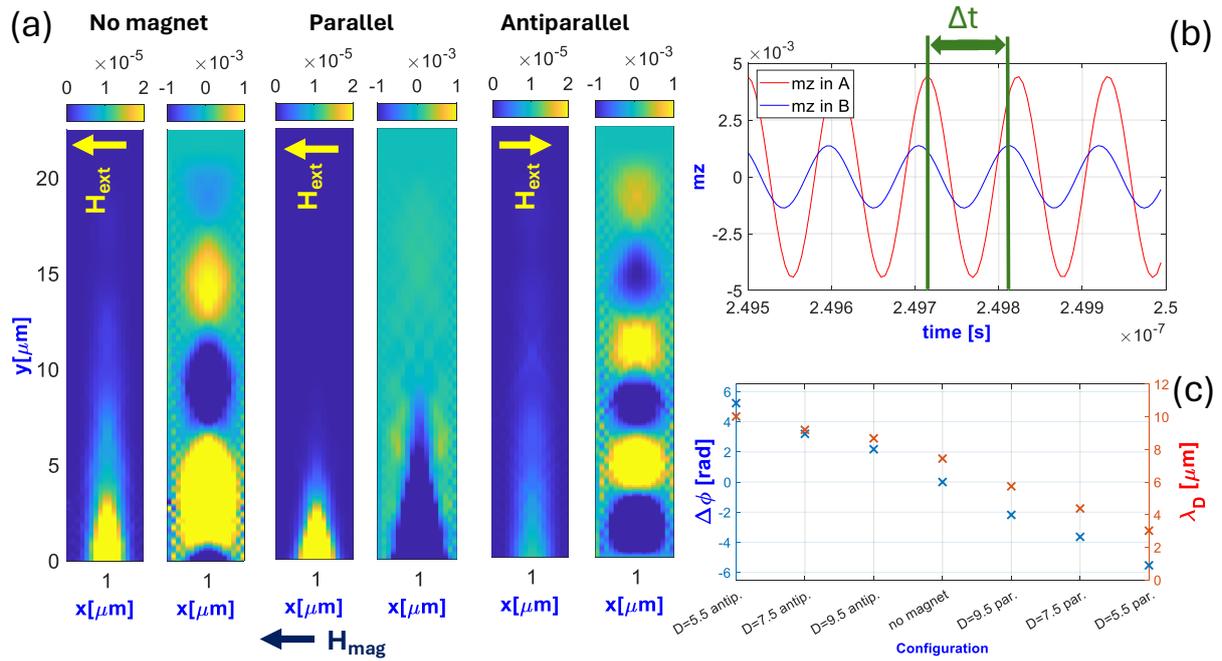

**Figure 4** a) Simulated spin-wave intensity profiles (left panel) and waveforms (right panel) of the main mode in the three configurations: absence of magnets, antiparallel, and parallel configuration, with the SmCo magnet placed at D = 5.5 µm. b) Temporal variation of the z component of the magnetization at a point in the waveguide 1 µm away from the antenna (red line) and 13 µm away from the antenna (blue line), showing the time delay between the two. c) Spin-wave decay length and phase shift as a function of distance in the parallel and antiparallel configurations.

# Supporting Information

## Tuning magnonic devices with on-chip permanent micromagnets

M. Cocconcelli, S. Tacchi, Róbert Erdélyi, F. Maspero, A. Del Giacco, A. Plaza, O. Koplak, A. Cattoni, R. Silvani, M. Madami, Á. Papp, G. Csaba, Felix Kohl, Björn Heinz, Philipp Pirro, R. Bertacco

**Section S1. VSM loops of SmCo films**

The static magnetic properties of W/SmCo/W films annealed at 650°C have been investigated by vibrating sample magnetometry. A typical loop measured between +- 9T is reported in figure S1. The saturation magnetization is 0.5 T while the coercive field is about 5000 Oe.

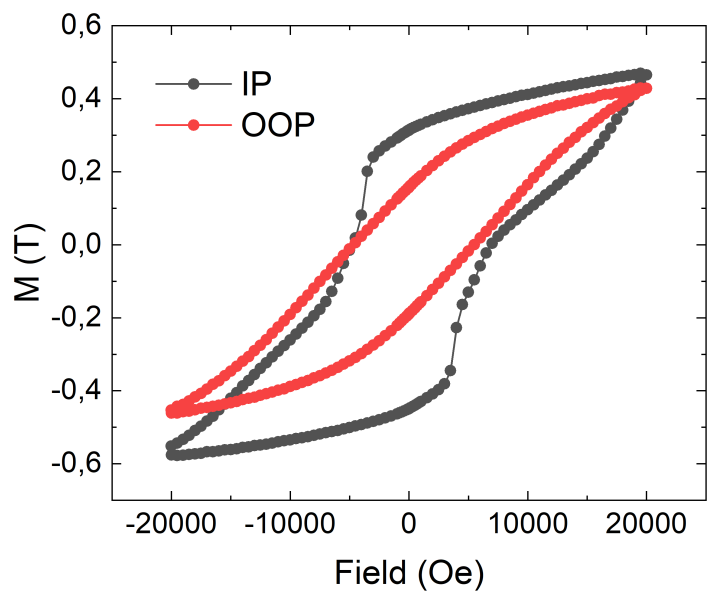

Figure S1: VSM loop from W(100)/SmCo(500)/W(100) films (thickness in nm).

**Section S2. CoFeB anisotropy characterization**

The properties of CoFeB films were investigated to determine important parameters that determine the spin wave propagation, such as the saturation magnetization, Gilbert damping and the eventual presence of anisotropies particularly examining variations induced by film growth on a planarization layer of silica. Four Ta(5 nm)/CoFeB(25 nm)/Ta(5 nm) films were deposited on bare silicon (100) and on sol-gel silica layers with thicknesses of 200 nm, 400 nm, and 560 nm, spun on silicon (100). The silica layer thickness was determined via ellipsometry. Vibrating sample magnetometry (VSM) experiments revealed uniaxial anisotropy in all samples. The film grown on bare Si exhibited an in-plane anisotropy easy axis at 22° relative to the (110) direction, while samples with SiO2 showed anisotropies at slightly different angles (61°, 45°, and 47° for samples grown on 200 nm, 400 nm, and 560 nm, respectively). All the samples show comparable saturation magnetization at $1.2\times10^6$ A/m. In Figure 1f we show the case of CoFeB on bare Si (red line) and on Si planarized with 350 nm of sol-gel silica, corresponding to the case of the final device.

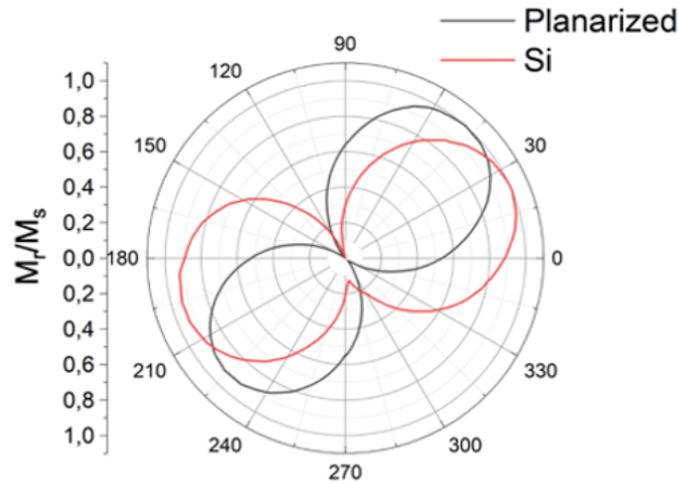

**Figure S2**: Polar plot of the ratio between the remanent and saturation magnetization of CoFeB film grown on bare silicon (red curve) and on a 400 nm thick planarization layer of silica (black curve) as a function of the angle θ between the saturation field and the [110] direction of Si.

To quantitatively estimate the films' anisotropy constants, vector network analyzer ferromagnetic resonance (VNA-FMR) experiments were conducted. Resonance frequency values were determined as a function of the angle by changing the relative orientation between the sample and the applied magnetic field. Results mirrored VSM findings, confirming uniaxial anisotropy in all samples, with angles similar to those found via magnetometry, as shown in **Figure S3**. Additionally, the anisotropy constant and Gilbert damping parameter were estimated for all films, as shown in **Table S1**.

The anisotropy value was employed to replicate the thermal spectra measured in micro-BLS experiments using mumax3 simulations, aiming to align the experimentally observed ferromagnetic frequency with the simulated counterpart. However, the experimentally determined anisotropy constant (-1.27e3 J/m^3) appears insufficient for accurate experimental-simulation alignment. Simulation results suggest a correct anisotropy constant of 1.25e4 J/m^3. This discrepancy may be attributed to the sample's non-negligible topography. Although the planarization layer reduces local surface roughness, overall sample flatness is not fully restored, leading to bulging near the magnet positions. Stress induced by the curvature in the silica, a softer material, may consequently increase the anisotropy constant of the CoFeB waveguide.

To validate this hypothesis, micro-BLS experiments were conducted on CoFeB waveguides of identical dimensions to those used in the study, patterned on bare Si. In this scenario, the experimentally observed ferromagnetic resonance was lower compared to the study's samples, with the thermal signal aligning perfectly with simulations using the anisotropy constant determined experimentally via VNA-FMR. This suggests that the presence of silica sol-gel and the topography induced by thick magnets contribute to the increased anisotropy constant due to stress.

Given this logical explanation for the increase, the anisotropy constant derived from the matching between simulation and micro-BLS experiment was utilized for all subsequent simulations.

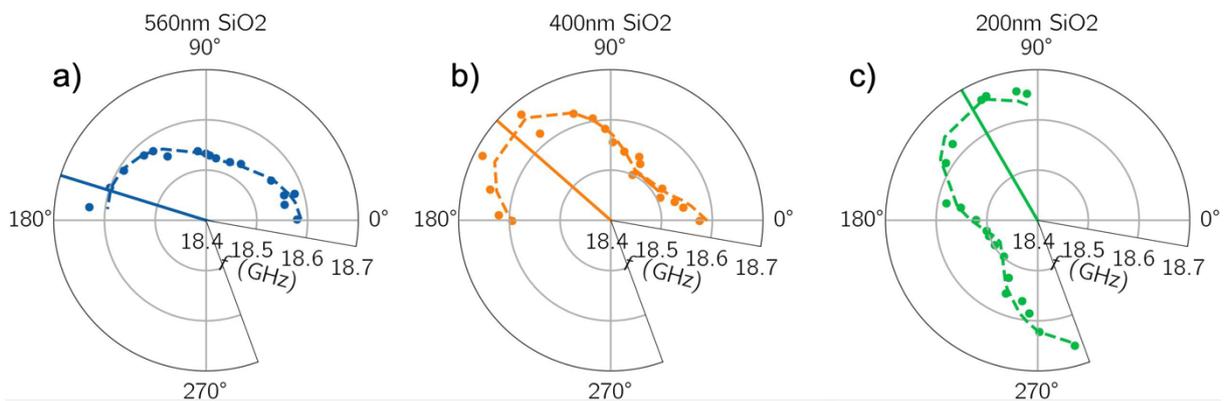

**Figure S3** Ferromagnetic resonance frequency plotted against the angle θ between the external magnetic field and the (110) Si crystallographic axis. The CoFeB films are deposited atop sol-gel layers with varying compositions, yielding SiO2 thicknesses of a) 560 nm, b) 400 nm, and c) 200 nm.

| Planarization layer | $K_u$ (J/$m^3$) | $M_s$ (kA/m) | $H_{K_u}$ (mT) | $\alpha_G$ ($10^{-3}$) |
|---|---|---|---|---|
| 560nm SiO$_2$ | -534 ± 64 | 1219 ± 0.5 | −0.9 ± 0.1 | 4.2 ± 0.1 |
| 400nm SiO$_2$ | -1204 ± 99 | 1222 ± 0.6 | −2.0 ± 0.2 | 5.4 ± 0.1 |
| 200nm SiO$_2$ | −1273 ± 74 | 1220 ± 0.5 | −2.1 ± 0.1 | 4.7 ± 0.1 |
| Si | −1303 ± 45 | 1221 ± 0.3 | −2.1 ± 0.1 | 5.3 ± 0.1 |

**Table S1:** Results of the investigation into the properties of CoFeB films as influenced by the type of planarization layer utilized.

## Section S3. Dispersion curves for CoFeB conduits under uniform external field

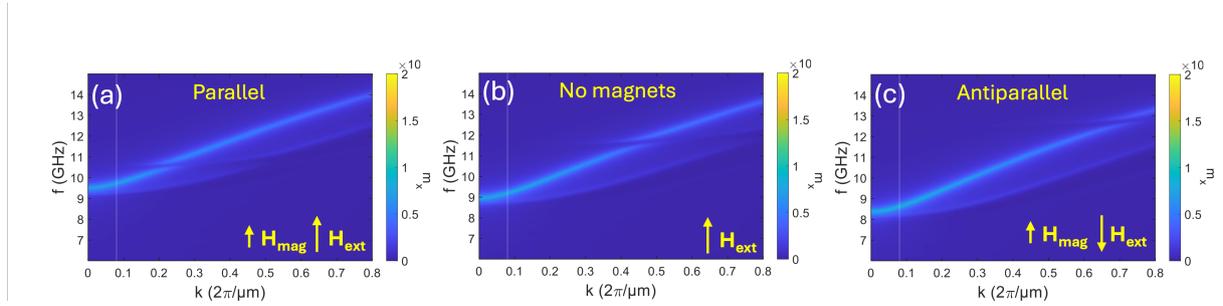

**Figure S4:** Dispersion plots from the 2D Fourier transformation of simulated signals are reported in panels a,b,c for the antiparallel, no-magnet, and parallel configuration corresponding to the experimental data on the device with SmCo magnets placed 5.5 µm from the centerline of the conduit.

## Section S4. Excitation frequency of the antenna and dispersion relations of the CoFeB conduit

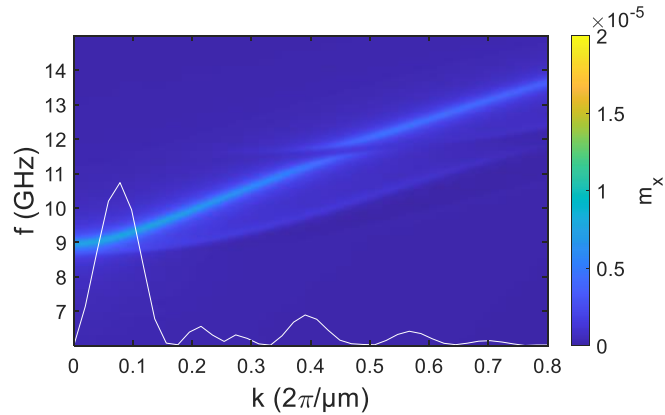

**Figure S5:** Excitation frequency of the CPW superposed to the band dispersion (no-magnets)

## Section S5. Layout of the simulated device

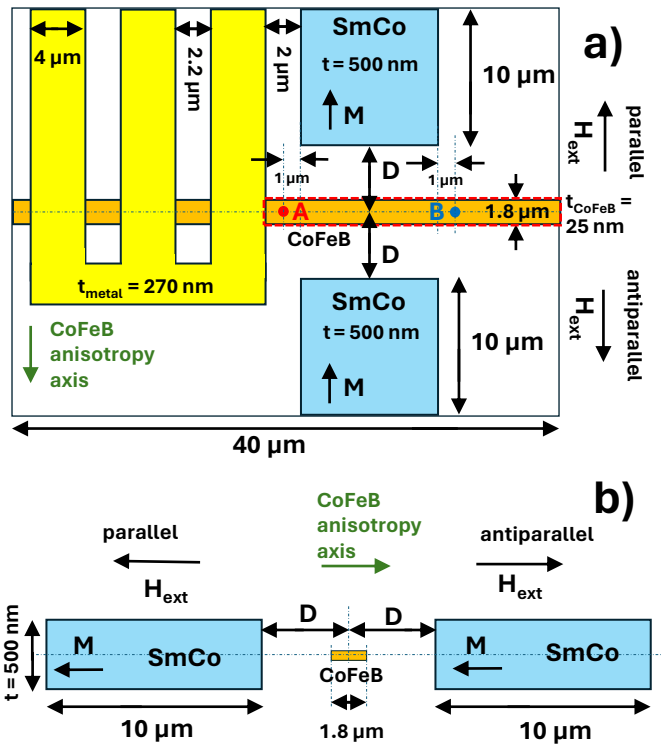

**Figure S6:** Sketch of the device layout used for simulations

## Section S6: Micro-BLS as a function of excitation frequency

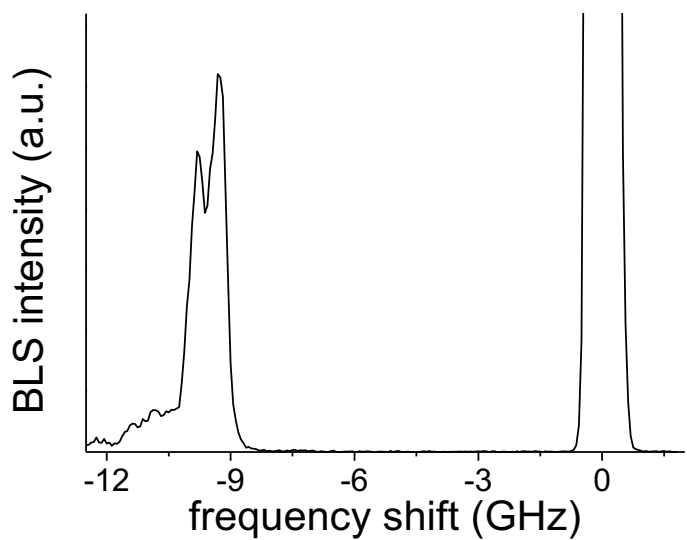

**References**

[1] S. S. Hamid, S. Mariappan, J. Rajendran, A. S. Rawat, N. A. Rhaffor, N. Kumar, A. Nathan, B. S. Yarman, **2023** *Micromachines* 14, 1551.

[2] P. Chen, G. Li, Z. Zhu, **2022** *Micromachines* 13, 656.

[3] M. D. Sinanis, P. Adhikari, T. R. Jones, M. Abdelfattah and D. Peroulis, **2022** IEEE Access 10, 19643.

[4] M. Abdelfattah and D. Peroulis, **2019** IEEE Trans. Microw. Theory Tech. 67, 3661.

[5] J. Chang, M. J. Holyoak, G. K. Kannell, M. Beacken, M. Imboden and D. J. Bishop, **2018** J. Microelectromechanical Syst. 27, 1135.

[6] C. H. Ko, K. M. J. Ho, R. Gaddi and G. M. Rebeiz, "A 1.5–2.4 GHz tunable 4-pole filter using commercial high-reliability 5-bit RF MEMS capacitors," 2013 IEEE MTT-S International Microwave Symposium Digest (MTT), Seattle, WA, USA, 2013, pp. 1-4.

[7] O. L. Balysheva, "Problems of Development of Tunable SAW Filters for Mobile Communication Systems," 2020 Wave Electronics and its Application in Information and Telecommunication Systems (WECONF), St. Petersburg, Russia, 2020, pp. 1-6.

[8] P. Pirro, V. I. Vasyuchka, A. A. Serga, B. Hillebrands, **2021** Nat. Rev. Mater. 6, 1114.

[9] V. G. Harris, **2012** IEEE Trans. Magn. 48, 1075.

[10] J. M. Owens, J. H. Collins and R. L. Carter, **1985** Circuits Syst. Signal Process. 4, 317.

[11] J. D. Adam, **1988** Proc. IEEE 76, 159.

[12] A. V. Chumak, P. Kabos et al., **2022** IEEE Trans. on Magnetics 58, 1.

[13] A. Barman et al. **2021** J. Phys.: Condens. Matter 33, 413001.

[14] Giacomo Talmelli et al., Sci. Adv.6, eabb4042(2020).

[15] B. Dieny, I. L. Prejbeanu, K. Garello, P. Gambardella, P. Freitas, R. Lehndorff, W. Raberg, U. Ebels, S. O. Demokritov, J. Akerman, A. Deac, P. Pirro, C. Adelmann, A. Anane, A. V. Chumak, A. Hirohata, S. Mangin, Sergio O. Valenzuela, M. Cengiz Onbaşlı, M. d'Aquino, G. Prenat, G. Finocchio, L. Lopez-Diaz, R. Chantrell, O. Chubykalo-Fesenko, P. Bortolotti, **2020** Nature Electron. 3, 446.

[16] K. Vogt, H. Schultheiss, S. Jain, J. E. Pearson, A. Hoffmann, S. D. Bader, B. Hillebrands, **2021** Appl. Phys. Lett. 101, 042410.

[17] K. Vogt, F.Y. Fradin, J.E. Pearson, T. Sebastian, S.D. Bader, B. Hillebrands, A. Hoffmann, H. Schultheiss **2014** Nat. Commun. 5, 3727.

[18] C.S. Davies, A. Francis, A.V. Sadovnikov, S.V. Chertopalov, M.T. Bryan, S.V. Grishin, D.A. Allwood, Y.P. Sharaevskii, S.A. Nikitov, and V.V. Kruglyak, **2015** Phys. Rev. B 92, 020408(R).

[19] A.V. Sadovnikov, C.S. Davies, V.V. Kruglyak, D.V. Romanenko, S.V. Grishin, E. N. Beginin, Y.P. Sharaevskii, S.A. Nikitov, **2017**, Phys. Rev. B 96, 060401(R).

[20] H. Qin, R.B. Holländer, L. Flajšman, S. van Dijken, Nano Lett. **2022**, 22, 5294; A. A. Serga, A. V. Chumak, B. Hillebrands, R. L. Stamps, M. P. Kostylev, **2010** EPL 90, 27003.

[21] G.Talmelli, D. Narducci, F. Vanderveken, M. Heyns, F. Irrera, I. Asselberghs, I. P. Radu, C. Adelmann, F. Ciubotaru, **2021** Appl. Phys. Lett. 118, 152410.

[22] O. Koplak, F. Maspero, A. Plaza, A. D. Giacco, M. Cocconcelli and R. Bertacco, **2023** IEEE Magn. Letters 14, 6100605.

[23] O. Koplak et al, submitted to Journal of Magnetism and Magnetic Materials

[24] V. E. Demidov, S. Urazhdin, A. B. Rinkevich, G. Reiss, and S. O. Demokritov, **2014** Appl. Phys. Lett. 104, 152402.

[25] A. I. Nikitchenko and N. A. Pertsev 2021 Phys. Rev. B 104, 134422.

[26] M. Mohseni, R. Verba, T. Brächer, Q. Wang, D. A. Bozhko, B. Hillebrands, and P. Pirro,**2019** Phys. Rev. Let. 122, 197201.

[27]V. E. Demidov, J. Jersch, S. O. Demokritov **2009** Phys. Rev. B 79, 054417